\documentclass[dvips]{acta}
\usepackage{supertabular,lscape,epsfig}
\usepackage{amssymb}
\usepackage{amsmath}
\usepackage[T1]{fontenc}

\SetPages{0}{0}

\SetVol{76}{2026}

\usepackage{lmodern}

\newcommand{\RES}{\Re(\sigma)}

\newcommand{\EKIN}{\mathcal{E}_{\rm kin}}

\newcommand{\NMC}{N_{\rm c}}
\newcommand{\NMH}{N_{\rm ha}}
\newcommand{\NMS}{N_{\rm sg}}
\newcommand{\NMI}{N_{\rm in}}

\newcommand{\NMR}{N_{\rm rsp}}
\newcommand{\CHIE}{\chi_{\rm eff}}
\newcommand{\LTEF}{\log(T_{\rm eff})}

\newcommand{\mFig}[1]{Fig.~\ref{fig:#1}}
\newcommand{\mEq}[1]{Eq.~(\ref{eq:#1})}

\begin{document}

\begin{Titlepage}

\Title{Power balance and anomalous exchange in post-AGB pulsations}

\Author{Zalewski, J.}
{Independent researcher \\
e-mail: jan.zalewski.a2@gmail.com}

\end{Titlepage}

\Abstract{
We develop a complex balance relation for stellar pulsations whose real and imaginary parts give paired relations for the excitation rate and oscillation frequency. The formulation includes thermodynamic exchange, boundary power, and the inertial, compression, horizontal-area, and gravitational-stratification terms, without assuming weak nonadiabaticity. Applied to portions of an envelope, it provides cumulative diagnostics evaluated using the eigenfrequency and eigenfunctions of the complete pulsation problem. We show that a negligible global boundary contribution can coexist with substantial internal boundary power, which may largely compensate the thermodynamic exchange within the envelope. Applications to radial pulsations of post-AGB envelopes illustrate this behaviour for ordinary, strange, and anomalous-exchange modes. The latter have a negative effective norm and opposite signs of net exchange power and excitation rate. For a selected excited anomalous mode, the principal exchange occurs between the He II ionization region and the Z-bump, with less participation from the outer ionization layers than in the comparison modes. Classification by the largest response norm reveals distinct domains across the examined pulsation spectrum. The selected anomalous branch retains horizontal-area dominance across its transition from damping to excitation.
}
{stars: oscillations, stars: AGB and post-AGB, stars: interiors, instabilities, methods: analytical, methods: numerical}

\section{Introduction}

Integral relations provide a means of examining the connection between thermodynamic exchange and stellar pulsation. The work integral is commonly used to identify regions contributing to mode driving or damping and to relate the net exchange to the excitation rate. Relations involving the complex pulsation frequency also connect nonadiabatic exchange with the oscillation frequency, as discussed by Cox (1980), Aikawa (1985), and Christensen-Dalsgaard (2014). For strongly nonadiabatic modes, examining these relations requires accounting for the response of the envelope and for the boundary contributions.

In Zalewski (2026c) a balance relation for radial pulsations was derived from a pressure-volume-rate sesquilinear form. Its extension to nonradial pulsations was presented in Zalewski (2026d). The resulting balance contains, in addition to the inertial contribution, response terms associated with compression, horizontal-area deformation, and gravitational stratification. In Zalewski (2026e) their combined effect was expressed through an effective norm factor $\chi_{\rm eff}$, which relates the net exchange power to the excitation rate. This factor can become negative, reversing the usual sign relation: a mode may grow while its net exchange power is negative, or decay while that power is positive. When the global boundary contribution is small, the same reversal applies to the thermodynamic exchange power. We refer to modes with $\chi_{\rm eff}<0$ as anomalous-exchange modes.

These global properties raise the question of how the balance is satisfied within the envelope. A cumulative thermodynamic work integral describes only one contribution to the balance of a region extending from an interior point to the surface. Such a region also exchanges mechanical power with the rest of the envelope through its inner boundary. A small boundary contribution for the complete envelope does not imply that this contribution is small at interior cuts. Its inclusion is therefore necessary when comparing thermodynamic exchange with the response terms and the change in kinetic energy within a part of the envelope.

Here we retain the full complex balance underlying the preceding studies. Its real part recovers the excitation balance, while its imaginary part gives a corresponding balance for the oscillation frequency. We formulate cumulative diagnostics for both balances and examine their constituent terms using the eigenfrequency and eigenfunctions of the complete pulsation problem. Examples of ordinary and strange radial modes illustrate the role of internal boundary power. We then examine the spatial balance of a selected excited anomalous mode in a post-AGB envelope and compare it with those of ordinary and strange modes. Finally, we classify modes along the examined sequence by their dominant response term, placing the anomalous modes within the wider pulsation spectrum.

\section{Balance relation for excitation rate and oscillation frequency}
In our previous work (Zalewski 2026c,d,e) we have obtained a quadratic balance relation by taking the real part of a pressure volume-rate sesquilinear form. This yields a relation involving the mode growth rate. Here we retain the full complex pairing and examine its imaginary part to obtain the corresponding relation for the oscillation frequency. The derivation retains the boundary terms and does not assume weak nonadiabaticity.

\subsection{The complex pressure-volume balance}
\label{sec:complex_balance_start}

We adopt the pulsation variables, normalization, and equilibrium
coefficients of Zalewski (2026c,d,e). In particular, $p$ and $s$
are the dimensionless Lagrangian pressure and entropy perturbations,
$d$ is the relative radial displacement, and
$\sigma=\gamma+i\nu$ is the dimensionless complex eigenfrequency.
An overbar denotes complex conjugation. The independent variable
is $x=\ln(r/R_\odot)$, with $x_b$ and $x_s$ denoting the inner and
outer boundaries. All integrals below extend over this domain.

The balance relations of the preceding papers were obtained from
the pressure--volume-rate pairing
\begin{equation}
	\mathcal K_s=-p\,\overline{\sigma q},
	\qquad
	\mathcal W_s=\int_{x_b}^{x_s}C\mathcal K_s\,dx,
	\label{eq:pairing_start}
\end{equation}
where $q$ is the relative Lagrangian volume perturbation and $C$
is the real weight defined in the preceding papers. The real part
of $\mathcal W_s$ gives the work integral.

Using integration by parts and momentum equations, and
retaining the full boundary contribution, gives
\begin{equation}
	\mathcal W_s+\overline{\sigma}\mathcal{D}
	=\sigma N_{\rm in}
	+\overline{\sigma}(N_{\rm ha}+N_{\rm sg}).
	\label{eq:mechanical_balance_start}
\end{equation}
The norm terms are real and retain their definitions as given in
Zalewski (2026d,e). They describe the inertial, horizontal area,
and gravitational stratification contributions, respectively.
For radial pulsations,
\begin{equation}
	\begin{aligned}
		N_{\rm in}
		&=-|\sigma|^2\int_{x_b}^{x_s}CA_3A_2|d|^2\,dx,\\
		N_{\rm ha}
		&=4\int_{x_b}^{x_s}CA_3|d|^2\,dx,
		\qquad N_{\rm sg}=0,\\
		\mathcal{D}&=\left[Cp\,\overline d\right]_{x_b}^{x_s}.
	\end{aligned}
	\label{eq:radial_terms_start}
\end{equation}
For nonradial pulsations the same balance uses the corresponding
nonradial norms and the complete boundary contribution, including
the gravitational boundary term from Zalewski (2026d).
Thus $\overline{\sigma}\mathcal{D}$ in
Eq.~(\ref{eq:mechanical_balance_start}) denotes the full complex
boundary contribution, not only its pressure part.

Substituting $q=-A_4p-A_7s$ into the integral for
$\mathcal W_{\rm s}$ separates the compression and entropy
contributions, giving:
\begin{equation}
	N_{\rm c}=-\int_{x_b}^{x_s}CA_4|p|^2\,dx,
	\qquad
	\mathcal T=\int_{x_b}^{x_s}CA_7p\,\overline s\,dx,
	\qquad
	\mathcal W_s=\overline{\sigma}(-N_{\rm c}+\mathcal T).
	\label{eq:thermodynamic_split_start}
\end{equation}
When these quantities are substituted into \mEq{mechanical_balance_start} and rearranged it is obtained that
\begin{equation}
		\mathcal W\equiv-\overline{\sigma}(\mathcal{T}+\mathcal{D})
		=-\sigma N_{\rm in}
		-\overline{\sigma}(N_{\rm c}+N_{\rm ha}+N_{\rm sg}).
	\label{eq:complex_exchange_balance_start}
\end{equation}
Here $\mathcal W$ combines the complex thermodynamic exchange
power and the boundary contribution. Using complex exchange power $\mathcal{P}_{\rm ex}=-\overline{\sigma}\mathcal{T}$ and noting that $-\Re(\overline{\sigma}\mathcal{D})=\Delta B$ the balance relation of Zalewski (2026e) is recovered as 
\[
\Re\mathcal{W}=\Re\mathcal{P}_{\rm ex}+\Delta B.
\]
Equation~(\ref{eq:complex_exchange_balance_start}) is the common
starting point for the excitation and oscillation balances
considered below.

\subsection{Global relation for eigenfrequency}
The inertia norm $N_{\rm in}$ can be expressed in terms of kinetic energy as $2 \mathcal{E}_{\rm kin}=-N_{\rm in}$ (see Zalewski 2026d). Substituting kinetic energy into \mEq{complex_exchange_balance_start} it is obtained that
\begin{equation}
\mathcal{W}=2\sigma \mathcal{E}_{\rm kin}-\overline{\sigma}(N_{\rm c}+N_{\rm ha}+N_{\rm sg}).
\label{eq:baseWEkin}
\end{equation}

The norm terms can be combined into the response norm $N_{\rm rsp}=\NMC+\NMH+\NMS$.  In Zalewski (2026d) a response power was introduced $\mathcal{P}_{\rm rsp}=\gamma N_{\rm rsp}$. Here we generalize it to a complex response power as 
\[
\mathcal{W}_{\rm rsp}=\overline{\sigma} N_{\rm rsp}.
\]
Using the response power the \mEq{baseWEkin} may be rewritten as
\begin{equation}
	\mathcal{W}=2\sigma \mathcal{E}_{\rm kin}-\mathcal{W}_{\rm rsp}.
\label{eq:baseWWrsp}	
\end{equation}
\mEq{baseWWrsp} relates the net complex exchange power to the complex inertial and response powers.

The \mEq{baseWWrsp} can be written in a form similar to that derived by Cox (1980) - Eq.(9.43) by introducing variable $F=\mathcal{T}+\mathcal{D}$ and expressing $\mathcal{W}=-\overline{\sigma}F$.
\begin{equation}
	\overline{\sigma}F=\sigma \NMI +\overline{\sigma}\NMR.
\end{equation}
Since $\NMI=-|\sigma|^2\mathcal{I}$, where $\mathcal{I}$ is mode inertia, then for $\sigma\neq 0$ it is obtained that
\begin{equation}
	F=-\sigma^2 \mathcal{I}+\NMR,
\end{equation}
and hence
\begin{equation}
	\sigma^2=\frac{\NMR}{\mathcal{I}}-\frac{F}{\mathcal{I}}.
\label{eq:SigmaEQ}		
\end{equation}
In the notation used here the balance relation \mEq{SigmaEQ} contains square of complex pulsation frequency, thus a formal quadratic relation, however both terms on the right hand side are implicit functions of $\sigma$. Thus it is not a polynomial but a complex integral balance equation.

Using $\omega=-i\Omega_0\sigma$ and making the formal substitutions $\Sigma^2=-\Omega_0^2\NMR/\mathcal{I}$ and $C/J=\Omega_0^3\sigma F/\mathcal{I}$, multiplication of \mEq{SigmaEQ} by $i\omega$ after conversion to $\omega$ yields
\[
i\omega(\omega^2-\Sigma^2)=C/J.
\]
This is the algebraic form of Eq.~(9.43) of Cox (1980). But it does not imply that this equation and our \mEq{baseWWrsp} are identical, as they are obtained under different assumptions. In particular, our expression retains the boundary contribution $\mathcal{D}$, whereas boundary terms are neglected in Cox's equation.

Relations for the complex pulsation frequency were also derived by Aika\-wa (1985) and Christensen-Dalsgaard (2014). Aikawa introduced nonadiabatic weight functions to examine how different regions of the envelope contribute to the pulsation period, comparing them with their adiabatic counterparts. Here we develop the real and imaginary parts of the complex balance as paired diagnostics of the contributions within the envelope.

\subsection{Growth rate and oscillation frequency}
\label{app:GMNU}

Rearranging \mEq{baseWWrsp} as
\begin{equation}
	2\sigma \EKIN=\mathcal{W}_{\rm rsp}+\mathcal{W}
\label{eq:baseEKIN},
\end{equation}
the complex inertial power is expressed as the sum of the exchange power including surface terms and the response power, and since both $\mathcal{E}_{\rm kin}$ and $N_{\rm rsp}$ are real then taking the real and imaginary parts of \mEq{baseEKIN} it is obtained that
\begin{equation}
		2\gamma \EKIN= \phantom{-}\gamma \NMR +\Re{\mathcal{W}},
	\label{eq:baseEKINRe}	
\end{equation}
\begin{equation}
		2 \nu \EKIN = -\nu\NMR +\Im{\mathcal{W}}.
	\label{eq:baseEKINIm}	
\end{equation}

These equations show that the excitation rate, \mEq{baseEKINRe}, and the oscillation frequency, \mEq{baseEKINIm}, are affected by the same factors - the total and the response powers and mode kinetic energy. 

Thus the relation expressed by \mEq{baseEKINRe} is paralleled by a symmetric form of the balance relation for the oscillation frequency. Both of these relations can be used for the same purposes - to examine which regions in the envelope are contributing most, or as a check of the consistency of numeric integration of pulsation equations. 

\subsection{Integral consistency checks}

When \mEq{baseEKINRe} and \mEq{baseEKINIm} are evaluated globally, using eigensolutions of the pulsation problem they represent identities. Based on that a check of the consistency of solution of the linear problem may be obtained. It is customary to rearrange \mEq{baseEKINRe} as
\begin{equation}
\gamma_{check}=\frac{\gamma \NMR+\Re\mathcal{W}}{2 \EKIN}
\label{eq:CheckRe}
\end{equation}
and evaluate the right hand side using the computed eigenfrequency and eigenmode obtaining the quantity $\gamma_{check}$. The value computed in this way should agree with the real part of the eigenfrequency - $\gamma$ and the difference may indicate deterioration of integration of the linear equations, for example due to the equations becoming stiff or other numeric problems.

While it is not common to use an equation for the checking of accuracy of the oscillation frequency, the \mEq{baseEKINIm} enables introduction of such a check
\begin{equation}
\nu_{check}=\frac{-\nu \NMR+\Im\mathcal{W}}{2 \EKIN}.
\label{eq:CheckIm}
\end{equation}

These comparisons test the consistency of the computed eigenfunction, eigenfrequency and numerical evaluation of the balance integrals.

\subsection{Limiting cases}
Using \mEq{baseWWrsp} and taking its real and imaginary parts it is obtained that
\begin{equation}
	\Re\mathcal{W}=\gamma\left(2\mathcal{E}_{\rm kin}-\NMR\right), \qquad
	\Im\mathcal{W}=\nu\left(2\mathcal{E}_{\rm kin}+\NMR\right).
\label{eq:WWrspReIm}
\end{equation}
	
	For a vanishing $\mathcal{W}=0$, which may occur when both the thermodynamic term $\mathcal{T}$ and boundary term $\mathcal{D}$ vanish or cancel out two types of modes are possible. A pure oscillation with $\gamma=0, \nu\neq 0$ for which case $\NMR=-2\EKIN$, or a pure growth or decay $\nu=0, \gamma\neq 0$ for which case $\NMR=2\EKIN$.  
	
	If, however a mode has both $\gamma\nu \neq 0$ then $\mathcal{W}\neq 0$. Thus an oscillatory excited or damped mode may exist only when the net complex exchange including boundaries is nonzero. It also means that if the thermodynamic term vanishes, an oscillatory (growing or decaying) mode requires a nonzero boundary contribution. This shows that the selection of the form of boundary conditions should be made taking into account their impact.

\subsection{Effective norm factor}
The \mEq{WWrspReIm} may be written using the effective norm factor $\CHIE$ (Zalewski 2026e)
\[
\CHIE=1-\frac{\NMR}{2\EKIN}.
\]
Using $\CHIE$ these equations are written as
\[
\Re\mathcal{W}=2\gamma\EKIN\CHIE, \qquad \Im\mathcal{W}=2\nu\EKIN(2-\CHIE).
\]

Neglecting $\NMR$ gives $\CHIE=1$, and the real part of the balance reduces to $\Re\mathcal{W}=2\gamma\EKIN$. For the balance relation derived in Zalewski (2026c) and discussed for radial pulsations in post-AGB envelopes (Zalewski 2026e) it was found that not only the $\CHIE$ magnitude but also its sign depend on the type of mode, see discussion in section~\ref{app:ThreeSpectrum}

For an adiabatic ($s=0$) oscillatory ($\nu\neq 0$) mode with vanishing boundary term $\mathcal{D}=0$ it follows from $\Im\mathcal{W}=0$ that $\CHIE=2$. Thus for adiabatic pulsation the value of the effective norm factor is $2$. 

\subsection{Spatial decomposition of the balance relation}

Except for global checks of the consistency of computations given by \mEq{CheckRe} and \mEq{CheckIm} the equations that follow from \mEq{baseEKIN} enable to obtain cumulative formulae describing the terms entering the balance relation and their dependence on location in the envelope. Such analyses are usually performed to identify terms in the balance equation and its derived forms that contribute to mode excitation (see eg. Zalewski 2026d). With the present approach these analyses can be extended to include also factors affecting the oscillation frequency.

By introducing cumulative norms in the following way
\begin{equation}
	\NMC(x) = -\int_{x}^{x_s}CA_4|p|^2\,dx,
\end{equation}
and similarly for the other response terms and kinetic energy and for cumulative power and difference of surface terms as
\begin{equation}
	\begin{aligned}
		\mathcal{T}(x) &= \int_{x}^{x_s} CA_7p\overline{s}\,dx,\\
		\mathcal{D}(x) &= \left[Cp\overline{d}\right]_{x}^{x_s},
	\end{aligned}
\end{equation}
and defining $\mathcal{E}_{\rm kin,b}=\mathcal{E}_{\rm kin}(x=x_b)>0$ the balance relation may be obtained for a region of the envelope bounded by its surface ($x_s$) and the current $x$. The balance relation may be computed on such a subrange of the pulsation region using the computed eigenfrequency and eigensolutions of the linear boundary value problem. Thus, for example the complex exchange power $\mathcal{P}_{\rm ex}(x)=-\overline{\sigma} \mathcal{T}(x)$ may be followed for $x$ spanning the pulsation region limits $x_b\leq x\leq x_s$ to obtain information where in the envelope the real part of the cumulative exchange power increases or decreases thus inferring about the mode driving regions. 

\begin{figure}[htb]
	\includegraphics{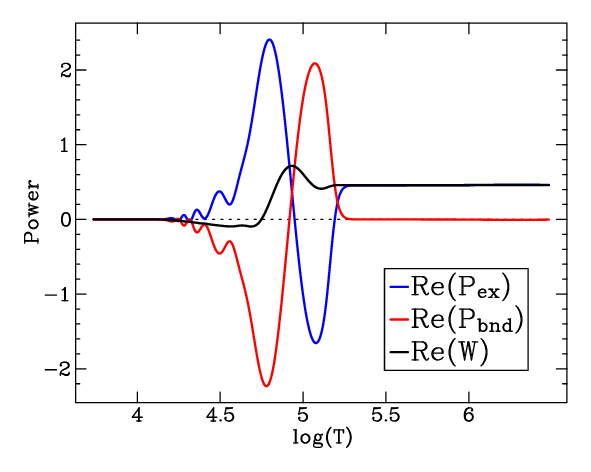}
	\FigCap{The dependence of real parts of cumulative exchange power $\Re\mathcal{P}_{\rm ex}(t)$, boundary-power $\Re\mathcal{P}_{\rm bnd}(t)$ and the total exchange power including surface terms $\Re\mathcal{W}(t)$ are shown as a function of $t=\log(T)$ for the 8-th overtone p-mode for $\LTEF=3.8$. The quantities shown are normalized by twice the kinetic energy at the bottom $2\EKIN(x_b)$. It is seen that in the region of strong compensation the exchange and pressure-work terms are of large magnitudes, much larger than their sum $\Re\mathcal{W}$ and are of opposite signs for this mode. Thus within the envelope the pressure-work term need not be small.   }
	\label{fig:Fig1}
\end{figure}  

With our formulation of the balance relation the complex boundary-power contribution $\mathcal{P}_{\rm bnd}=-\overline{\sigma} \mathcal{D}(x)$ requires closer examination. The global boundary-power contribution, when it appears in the global balance relation, is obtained from the difference of terms computed at both boundaries, and with a suitable choice of boundary conditions may be made much smaller than the $\mathcal{P}_{\rm ex}$ term (see Zalewski 2026e) - when the boundary conditions are selected so as not to cause disturbance in power balance by forcing the eigensolutions near boundaries. However when one considers a subregion of the pulsation region bound by $[x,x_s]$ then it turns out that the $\mathcal{P}_{\rm bnd}(x)$ need not be small. In fact since the real part of the $\mathcal{P}_{\rm bnd}(x)$ for radial pulsation may be interpreted as the difference of acoustic flux between the outer and inner boundaries (pressure-work) then it may be expected that in regions where the pressure and displacement perturbations are large the $\mathcal{P}_{\rm bnd}$ may attain large values.

This may be seen in \mFig{Fig1} where the real parts of $\mathcal{P}_{\rm ex}(t), \, t=\log(T)$ as well as $\mathcal{P}_{\rm bnd}(t)$ and their sum $\mathcal{W}(t)$ are plotted for 8-th overtone p-mode for $\LTEF=3.8$ envelope. 

From \mFig{Fig1} it follows that for the considered mode the cumulative exchange power term $\Re\mathcal{P}_{\rm ex}$ starts to increase below He~II ionization zone followed by a decrease in the regions above the Z-bump and stabilizes on a positive value deep in the envelope. The cumulative pressure-work term $\Re\mathcal{P}_{\rm bnd}$ follows the behavior of the exchange term with opposite sign. Deep in the envelope it stabilizes to values close to zero. Thus while the global $|\Re\mathcal{P}_{\rm bnd}|\ll |\Re\mathcal{W}|$, locally, in the regions where substantial exchange occurs for this mode the pressure-work term is of comparable magnitude to the thermodynamic exchange power term. In the region of large exchange of power the net power $\Re\mathcal{W}(t)$ has smaller magnitude than any of the constituent terms. It stabilizes deep in the envelope becoming close to the exchange term since the pressure-work term becomes small.

\begin{figure}[htb]
	\includegraphics{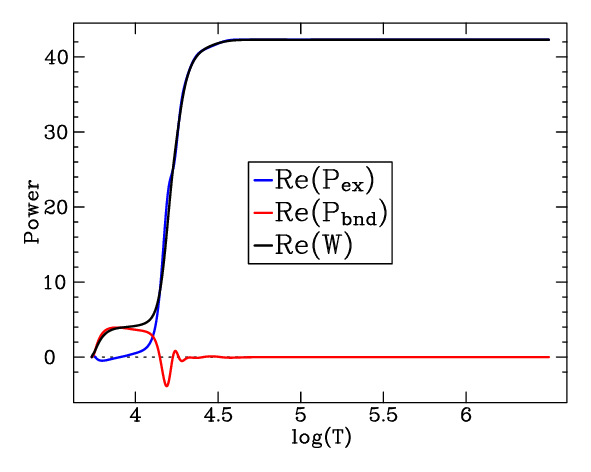}
	\FigCap{The dependence of real parts of cumulative exchange power $\Re\mathcal{P}_{\rm ex}(t)$, the boundary-power $\Re\mathcal{P}_{\rm bnd}(t)$ and the total exchange power including surface terms $\Re\mathcal{W}(t)$ are shown as a function of $t=\log(T)$ for the $S_3^+$ excited strange mode for $\LTEF=3.8$. The quantities shown are normalized by twice the kinetic energy at the bottom $2\EKIN(x_b)$. The boundary-work term for the strange mode plays a much smaller role in the power balance $\Re\mathcal{W}(t)$ than for ordinary mode. }
	\label{fig:Fig2}
\end{figure}  

In \mFig{Fig2} the same quantities are shown but for a $S_3^+$ excited strange mode. It may be seen that in this case the pressure-work term is also non-negligible in the driving region and becomes insignificant deep in the envelope. However for a strange mode the $\Re\mathcal{P}_{\rm bnd}$ has a substantially smaller magnitude than the exchange term and does not follow it with opposite sign as it was for the 8-th overtone ordinary mode. 

By examining the terms entering the net power $\mathcal{W}$ for ordinary and strange mode it is seen that the boundary-power term inside the envelope is not negligible and needs to be included in the proper balance relation. Therefore it is not only the thermodynamical term $\Re\mathcal{P}_{\rm ex}$, dependent on the $p\overline{\sigma s}$ sesquilinear form, that needs to be considered when analyzing mode driving. The net power $\Re\mathcal{W}(x)$ can differ substantially from the thermodynamic contribution alone.

\subsection{Normalized cumulative balance contributions}
Using the quantities introduced in the preceding section and \mEq{baseEKINRe} and \mEq{baseEKINIm} the cumulative factors $G$ may be defined as
\begin{equation}
	\begin{aligned}
		G_{\gamma}(x)&=\frac{\phantom{-}\gamma \NMR(x)+\Re(\mathcal{W}(x))}{2 \mathcal{E}_{\rm kin,b}}, \\
		G_{\nu}(x)   &=\frac{-\nu \NMR(x)+\Im(\mathcal{W}(x))}{2 \mathcal{E}_{\rm kin,b}}.
	\end{aligned}
\label{eq:EQG}
\end{equation}

The \mEq{EQG} represents the right hand side of the balance relations normalized by twice the total kinetic energy. The normalization is introduced so that the results can be compared among modes and models.

The quantities $G(x)$ are the kinetic energy weighted real and imaginary parts of the frequency - as given by the balance relation extended from surface to the given point $x$ in the envelope:
\[
\begin{aligned}
G_{\gamma}(x)&=\frac{\gamma \mathcal{E}_{\rm kin}(x)}{\mathcal{E}_{\rm kin,b}}, \\
G_{\nu}(x)&=\frac{\nu \mathcal{E}_{\rm kin}(x)}{\mathcal{E}_{\rm kin,b}},
\end{aligned}
\]
and provided $\nu\neq 0$ and $\EKIN(x)>0$ the ratio $G_\gamma(x)/G_\nu(x)=\gamma/\nu$ is constant. At the lower boundary
\[
\gamma =G_{\gamma}(x=x_b), \qquad \nu =G_{\nu}(x=x_b).
\] 

The expressions for $G_{\gamma}(x)$ and $G_{\nu}(x)$ do not define the real and imaginary parts of a spatially variable eigenfrequency $\sigma(x)$.
Instead they represent the eigenfrequency weighted by the fractional kinetic energy contained between surface and the particular layer of the envelope.
The constituent terms of both of these expressions show how the excitation or oscillation balances are satisfied within the envelope.
We shall examine their properties in the following section.

\section{Anomalous-exchange modes}
We refer to modes with $\CHIE<0$ found by Zalewski (2026e) as anomalous-exchange modes. For $\gamma\neq0$, the balance relation $\Re\mathcal{W}=2\gamma\EKIN\CHIE$ implies that the total exchange power $\Re\mathcal{W}=\Re\mathcal{P}_{\rm ex}+\Delta B$ and the growth rate $\gamma$ have opposite signs. Excited anomalous modes therefore have $\Re\mathcal{W}<0$, whereas damped anomalous modes have $\Re\mathcal{W}>0$. When the boundary contribution $\Delta B$ is sufficiently small to leave the sign of the total exchange unchanged, the thermodynamic exchange power $\Re\mathcal{P}_{\rm ex}$ also has the opposite sign to $\gamma$. Both behaviors were found in Zalewski (2026e), including an excited mode with negative thermodynamic exchange and a small boundary contribution. Excited oscillatory anomalous modes occur at higher effective temperatures, while damped anomalous modes include ordinary p-mode overtones at $\LTEF<4$. We focus below on a selected low-frequency excited anomalous mode and compare its spatial balance with those of ordinary and strange modes.

Here we examine radial pulsations along a sequence of post-AGB envelope models spanning $3.52\leq\LTEF\leq4.6$, with $L=10^4L_\odot$ and $L/M=14500\,L_\odot/M_\odot$. The pulsation calculations use boundary conditions of the form $(3,4)\texttt{-}(1,3)$, following the discussion of their effect on the spectrum, including strange modes, in Zalewski (2026a). The equations are integrated using continuous renormalization (Zalewski 2026b). For the present analysis, the survey of Zalewski (2026e) was supplemented by computations of the boundary-power and other balance contributions.

\begin{figure}[htb]
	\includegraphics{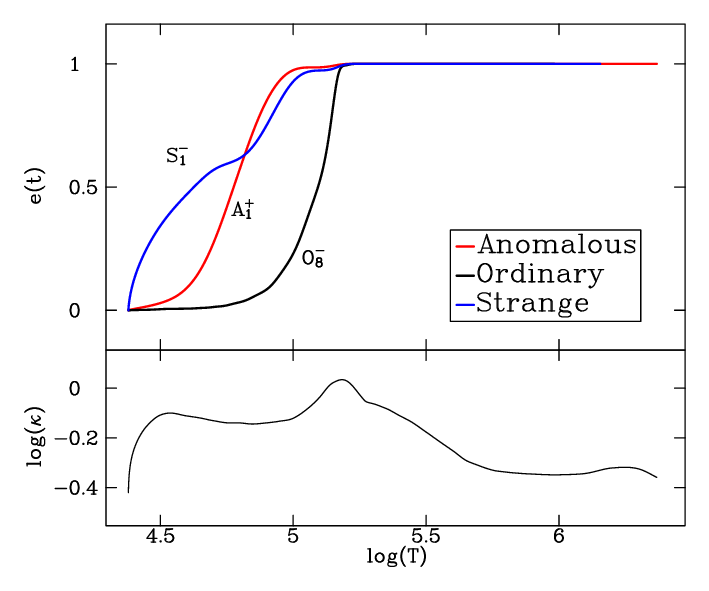}
	\FigCap{In the upper panel the cumulative kinetic energy integral profile normalized by the kinetic energy at the bottom boundary $e(t)=\mathcal{E}_{\rm kin}(t)/\mathcal{E}_{\rm kin,b}$ is shown for three modes: the anomalous - $A_1^+$, and ordinary damped overtone $O_8^-$ and a damped strange mode $S_1^-$ for effective temperature $\LTEF=4.45$. In the lower panel a $\log_{10}\kappa$ is given showing the location of the Z-bump. The He ionization occurs near surface. For the strange mode the cumulative kinetic energy increases rapidly near the surface and then near the Z-bump. For the ordinary mode $\EKIN(t)$ increases steeply at the onset of the Z-bump, while the kinetic energy of the anomalous mode  increases rapidly in the regions below He~II ionization and above the Z-bump.}
	\label{fig:Fig3}
\end{figure}  

The normalized kinetic energy dependence on $\log(T)$ in the envelope for three modes for the envelope model with $\LTEF=4.45$ are shown in \mFig{Fig3}. The dependence on position in the envelope of $e(t)=\mathcal{E}_{\rm kin}(t)/\mathcal{E}_{\rm kin,b}$, i.e. $G_{\gamma}(t)/\gamma$ is shown for three modes - viz. anomalous excited mode with $\sigma=(0.007, 0.386)$ denoted as $A_1^+$, an ordinary overtone p-mode with $\sigma=(-0.312, 5.812)$ denoted as $O_8^-$, and a damped strange mode with $\sigma=(-0.475, 1.533)$, denoted as $S_1^-$.

As may be seen for the anomalous mode $A_1^+$ the cumulative kinetic energy increases predominantly in the regions below the He~II and above the Z-bump. The strange mode $\mathcal{E}_{\rm kin}(t)/\mathcal{E}_{\rm kin,b}$ increase parallels that of the anomalous mode near the Z-bump but the main increase occurs close to the surface in the He ionization zones. For the ordinary mode most of the kinetic energy is concentrated near the onset of the Z-bump. Thus the three modes' dynamics is determined in different parts of the envelope.

\begin{figure}[htb]
	\includegraphics{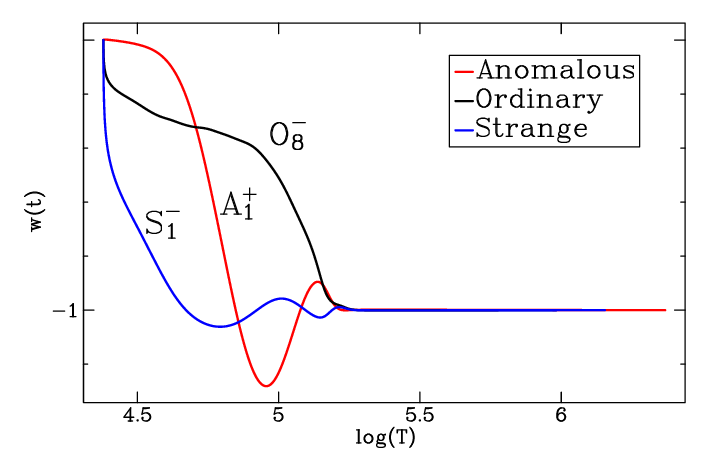}
	\FigCap{The real part of the cumulative work integral including boundary-power normalized by the absolute value of the work at the bottom boundary $w(t)=\Re\mathcal{W}(t)/|\Re\mathcal{W}(x_b)|$ is shown for three modes: the anomalous - $A_1^+$, and ordinary damped overtone $O_8^-$ and a damped strange mode $S_1^-$. For the strange mode the cumulative work decreases rapidly near surface. For the ordinary mode the work decreases also near surface, but the main decrease is near the Z-bump region. For the anomalous mode the large decrease of $w(t)$ occurs in the regions between He~II ionization zone and the Z-bump.}
	\label{fig:Fig4}
\end{figure}  

The real part of the normalized exchange power including the boundary-power, $w(t)=\Re\mathcal{W}(t)/|\Re\mathcal{W}(x_b)|$, is shown in \mFig{Fig4} for the same three modes as in \mFig{Fig3}. The work for each mode is normalized by the absolute value of $\Re\mathcal{W}$ at the bottom of the envelope to make the profile shapes comparable. The $\Re\mathcal{W}$ as normalized by mode's kinetic energy at the bottom boundary would be closest to zero for the anomalous mode, followed by the ordinary mode and largest (most dissipation) for the damped strange mode.

Similarly to the case with normalized cumulative kinetic energy most of the anomalous mode cumulative work occurs in the regions between the He~II ionization zone and the Z-bump region. However since the combined response power for this mode ($\Re\mathcal{P}_{\rm rsp}$) is positive and exceeds the $\Re\mathcal{W}$ in magnitude the overall excitation rate for the anomalous mode is positive, $\Re(\sigma)>0$. Modes with anomalous relation of the total work $\Re\mathcal{W}$ to the excitation rate $\gamma$ are characterized by negative effective norm factor $\CHIE<0$ as discussed in Zalewski (2026e). 

\begin{figure}[htb]
	\includegraphics{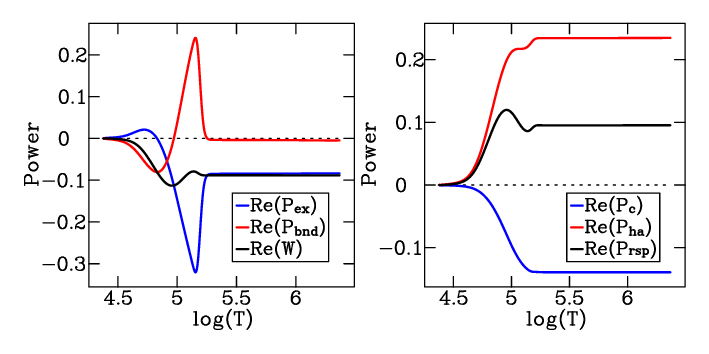}
	\FigCap{Left panel - the real part of cumulative exchange power $\Re\mathcal{P}_{\rm ex}(t)$, the boundary-power $\Re\mathcal{P}_{\rm bnd}(t)$ and their sum $\Re\mathcal{W}(t)$ are shown as a function of $t=\log(T)$ for the $A_1^+$ excited anomalous mode. The quantities shown are normalized by twice the kinetic energy at the bottom $2\EKIN(x_b)$. The boundary-work term for the anomalous mode plays an important role, and as for the ordinary mode it counteracts the exchange power in the region close to Z-bump. Deeper in the envelope the boundary-power becomes much smaller than the exchange power and $\Re\mathcal{W}\approx\Re\mathcal{P}_{\rm ex}$ deep in the envelope. Right panel - the components of the cumulative response power. Horizontal area deformation makes the response power positive. }
	\label{fig:Fig5}
\end{figure}  

For the anomalous mode $A_1^+$ the two components entering into $\Re\mathcal{W}$ behave similarly to the ordinary mode, i.e. they oppose each other so that the resulting total work near the Z-bump is smaller in magnitude than any of them. This may be seen in \mFig{Fig5}. For the anomalous mode the boundary-power term plays an important role in the region near the Z-bump, but below this region this term becomes negligible for the adopted boundary conditions of the form of $(3,4)-(1,3)$ (see the discussion in Zalewski 2026e). As can be seen from the plots of $\Re\mathcal{P}_{\rm ex}$ and $\Re\mathcal{W}$ the boundary-power affects $\mathcal{W}$ near $\log(T)\approx 5$, but deeper in the envelope the total exchange plus boundary power follows the exchange power. Thus the global boundary contribution is negligible and the mode is unstable ($\RES>0$) because the response power term $\Re\mathcal{P}_{\rm rsp}>|\Re\mathcal{W}|$. This is due to the horizontal area deformation term $\Re\mathcal{P}_{\rm ha}$ being large and positive and thus $2\gamma \mathcal{E}_{\rm kin}=\Re\mathcal{P}_{\rm rsp}+\Re\mathcal{W}>0$ even though $\Re\mathcal{W}\approx\Re\mathcal{P}_{\rm ex}<0$. Some of the properties of the anomalous mode are similar to ordinary modes (large and opposite exchange and boundary-power terms above Z-bump) and some to strange modes viz. near cancellation of exchange power and response power terms, but unlike strange modes (Zalewski 2026c) it is the horizontal area deformation component that is important. The anomalous mode $A_1^+$ displacement perturbation decreases inwards and in the Z-bump region is already small, while for an ordinary overtone modes it is this region where their amplitudes become large. Strange modes amplitudes also decrease below H/He ionization zones. The anomalous mode retains substantial amplitude higher than for strange modes down to regions close to the Z-bump.

\subsection{Dominant response terms across the spectrum}
\label{app:ThreeSpectrum}

The real part of the balance equation, as given by \mEq{baseEKINRe} may be rearranged by expressing the kinetic energy $\EKIN$ in terms of the inertia norm $\NMI$ as
\begin{equation}
\Re\mathcal{W}=-\gamma\left(\NMI+\NMC+\NMH+\NMS\right).
\label{eq:DistrEq}
\end{equation}
The inertia term may be treated as a response term. This equation represents how the exchange plus boundary-power $\Re\mathcal{W}$ is balanced by the four response terms. For radial pulsation, $\NMS=0$, leaving only three contributions. It is thus interesting to examine for which modes of pulsation these individual response terms dominate across the post-AGB range of effective temperatures. We have used the survey of pulsation modes from Zalewski (2026e) and for each mode determined which of the three terms makes the largest contribution using
\[
n_j=\frac{|N_j|}{\sum_i |N_i|},\qquad i,j \in \{\rm {in,c,ha}\}.
\]
This enables to classify modes into three types, according to the index for which the response term $n_j$ is largest. 

\begin{figure}[htb]
	\includegraphics{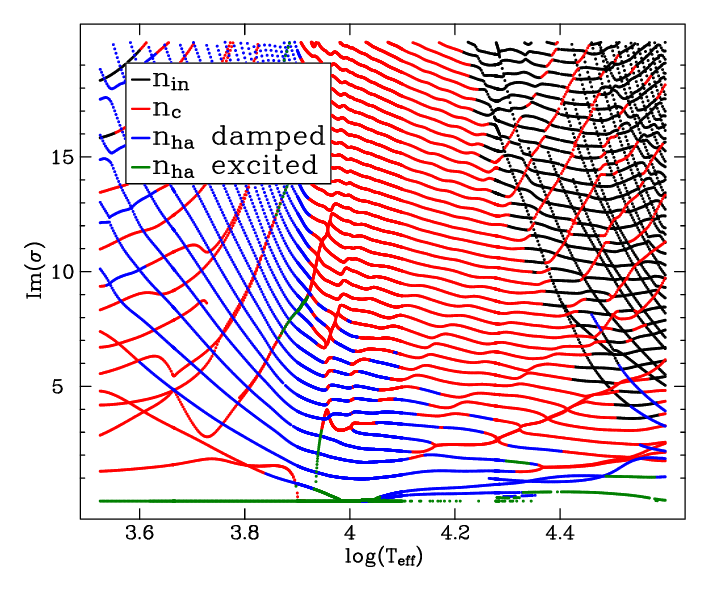}
	\FigCap{The spectrum of modes over a range of effective temperatures from $3.52\leq\LTEF\leq4.6$ for post-AGB sequence of models with $\log(L/L_{\odot})=4$. The black points - the $n_{\rm in}$ is largest, red points - the compression term ($n_{\rm c}$), and blue ones - the horizontal area deformation ($n_{\rm ha}$) are largest and the modes are damped, while green points - represent dominant $n_{\rm ha}$ and excited modes ($\RES>0$). Mode types according to dominant response term form well defined domains. }
	\label{fig:Fig6}
\end{figure}  

The results are shown in \mFig{Fig6}. In case of the horizontal area deformation type of modes ($n_{\rm ha}$ has largest value) the blue points represent modes which are damped while green points represent modes which are excited. For the discussion of strange modes see Zalewski (2026e).

From \mFig{Fig6} it may be seen that the domains of dominance of individual response terms in \mEq{DistrEq} are separated. At high temperatures ($\LTEF>4.3$) the sequences of ordinary modes as well as the sequences of modes with frequencies decreasing substantially with the increase of effective temperature (see Gautschy 1993) are both predominantly of the type for which the term $|N_{\rm in}|$ dominates, thus for these modes the inertia has largest contribution to the balance. In the central part ($4\leq\LTEF\leq 4.3$) there occur modes for which the compression term $|N_{\rm c}|$ is largest, also the low temperature strange modes as well as high temperature modes undergoing avoided crossings are of this kind. At lower temperatures ($\LTEF\lesssim 4$) the horizontal area deformation term $|N_{\rm ha}|$ dominates. This term is also important at higher temperatures for low frequency modes.

The region of $|N_{\rm ha}|$ dominance shown \mFig{Fig6} may be compared to the region marked as $\CHIE<0$ in Fig.~1 in Zalewski (2026e). It turns out that the region where the relation between supplied power and kinetic energy rate of change is reversed is enclosed in the region where the magnitude of the horizontal area deformation term dominates the response terms, since
\[
\CHIE<0 \iff |N_{\rm ha}|>|N_{\rm in}|+|N_{\rm c}|.
\] 

Besides the excited anomalous mode discussed above, two other excited anomalous modes of higher frequency are found near $\LTEF=4.3$ and $4.5$, respectively. Excited anomalous modes are also found at lower temperatures ($\LTEF<4$), where they are thermal rather than oscillatory (see \mFig{Fig6}).

For the studied anomalous mode $A_1^+$ near $\LTEF=4.45$ it may be seen from \mFig{Fig6} that it continues along the excited anomalous mode branch (green curve) to higher temperatures, while for lower effective temperatures it becomes damped anomalous mode and continues towards its origin near $\LTEF=4.06$. Thus this mode retains horizontal area deformation dominance along the sequence. But the higher frequency anomalous mode seen near $\LTEF=4.3$ becomes a compression-type mode following a crossing with another mode close to $\LTEF=4.4$.

\section{Conclusions}

We have introduced a single complex balance which yields paired relations for mode growth and oscillation frequency, using the same response norms and complex exchange plus boundary terms. Its real part recovers the power balance discussed in Zalewski (2026c,d,e).

The balance relation can be applied to a shell within the envelope and evaluated using the eigenfrequency and eigenfunctions of the complete boundary value problem. The boundary contribution must include both boundaries of that shell.

A negligible global boundary contribution does not imply negligible internal boundary work. At interior cuts, the boundary-power term $\Re\mathcal{P}_{\rm bnd}(x)$ can rival the exchange power $\Re\mathcal{P}_{\rm ex}(x)$, producing substantial compensation. The boundary-power term accounts for mechanical power exchanged between neighbouring regions, so the net power balance of a shell can differ substantially from its thermodynamic exchange alone, even when the global boundary contribution is negligible.

The cumulative quantities $G_\gamma(x)=\gamma E_{\rm kin}(x)/E_{\rm kin,b}$ and the similarly obtained $G_\nu(x)=\nu E_{\rm kin}(x)/E_{\rm kin,b}$ follow from the real and imaginary parts of the balance relation. Both depend on the cumulative kinetic energy, and their ratio (when defined) is $\gamma/\nu$ independently of $x$. Their components, however, contribute differently to the two balances. By examining these components it is possible to establish how the excitation and oscillation balances are satisfied within the envelope.

In Zalewski (2026e) it was found that positive mode growth can coexist with negative global thermodynamic exchange for the mode. The spatial decomposition of the mode's exchange, boundary-power and response contribution is essential to interpreting this coexistence. 

We have called modes for which $\CHIE<0$ anomalous-exchange modes. They occur within the domain where modes with the dominant horizontal area deformation term ($N_{\rm ha}$) are found. These modes occupy a large section of the spectrum diagram, as shown in \mFig{Fig6}. The excited anomalous modes appear at higher effective temperatures as low frequency modes, while at low temperatures as thermal modes. We have analyzed a  selected anomalous oscillatory mode and found that it differs spatially from the ordinary and strange modes (\mFig{Fig3}). For the anomalous mode the principal exchange is concentrated between the He~II ionization region and the Z-bump (\mFig{Fig4}), with less participation from the outer ionization layers than for the strange mode, and from the Z-bump than for the comparison ordinary mode.

Along the examined post-AGB sequence, modes with different dominant response terms occupy distinct domains in the spectrum. This classification complements the distinction between ordinary and strange modes. A further property is anomalous exchange: the excitation rate and the net exchange power $\Re\mathcal{W}$ have opposite signs.

\end{document}